\pdfoutput=1
\documentclass[aps,pre, twocolumn, groupedaddress]{revtex4-1}

\usepackage{lipsum}
\usepackage{mathtools}
\usepackage{graphicx}
\usepackage{dcolumn}
\usepackage{amsmath}    
\usepackage{amssymb}
\usepackage{bm}
\usepackage{hyperref}
\usepackage{latexsym}
\usepackage{verbatim}
\usepackage[normalem]{ulem}
\usepackage{color}
\usepackage[caption=false]{subfig}
\def\beq{\begin{equation}}
\def\eeq{\end{equation}}

\def\beq{\begin{equation}}                          
\def\eeq{\end{equation}}                          
\def\bea{\begin{eqnarray}}                          
\def\eea{\end{eqnarray}}

\DeclareRobustCommand{\uvec}[1]{{%
  \ifcsname uvec#1\endcsname
     \csname uvec#1\endcsname
   \else
    \bm{\hat{\mathbf{#1}}}%
   \fi
}}
\preprint{}
\begin{document}
\preprint{}
%%%%%%%%%%%%%%%%%%%%%%%%%%%%%%%%%%%%%%%%%%%%%%%%%%%
%                               TITLE & ABSTRACT
%%%%%%%%%%%%%%%%%%%%%%%%%%%%%%%%%%%%%%%%%%%%%%%%%%%
%Title of paper
\title{Bond disorder enhances the information transfer in the polar flock}
\author{Jay Prakash Singh$^{1}$}
\email{jayps.rs.phy16@itbhu.ac.in}
%\affiliation{Indian Institute of Technology (BHU) Varanasi, India 221005}
\author{Sameer Kumar$^{2}$}
\email{samkum2010@gmail.com}
%\affiliation{S. N. Bose National Centre for Basic Sciences, Kolkata, India 700106}
\author{Shradha Mishra$^{1}$}
\email{smishra.phy@iitbhu.ac.in}
\affiliation{$^{1,2}$Department of Physics, Indian Institute of Technology (BHU), Varanasi, India 221005}
%\affiliation{$^{2}$S. N. Bose National Centre for Basic Sciences, J D Block, Sector III, Salt Lake City, Kolkata 700106}
\date{\today}

\begin{abstract}
Collections of self-propelled particles (SPPs) exhibit coherent motion and show {\em true} long-range order in two-dimensions. Inhomogeneity, in general, destroys the usual long-range order of the polar SPPs. We model a system of polar self-propelled particles with inhomogeneous interaction strength or {\em bond disorder}. The system is studied near the order-to-disorder transition for different strengths of the disorder. Our numerical simulation indicates that the nature of the phase transition
changes from discontinuous to continuous type by tuning the strength of the disorder. The bond disorder also enhances the ordering near the transition due to the formation of a homogeneous flock state for the large disorder. 
It leads to faster information transfer in the system and enhances the systems' information entropy. Our study gives a new understanding of the effect of intrinsic inhomogeneity in the self-propelled particle system.
\end{abstract}
\maketitle
\section{Introduction \label{Introduction}}
{{Collective behavior of a large number of self-propelled particles (SPPs) or ``flocking'' is ubiquitous. Examples of such systems range from a few micrometers, e.g., actin and tubulin filaments, molecular motors \cite{Nedelec1997, Yokota1986,Toyo,kron,Laub}, unicellular organisms such as amoebae and bacteria \cite{Bonner1998}, to several meters, e.g., birds flock \cite{Chen2019}, fish school \cite{Parrish1997}, and human crowd \cite{Helbing2000}, \textit{etc}. Interestingly, these systems show a collective motion on a scale much larger than each individual and hence long-range ordering (LRO) is observed in two-dimensions.\\
A minimal model for understanding the basic features of the collective behavior of polar self-propelled  particles or  
 ``polar flock'' was introduced in 1995 by T. Vicsek {\em et al.} \cite{VicsekT}.  In the last three decades, many variants of the Vicsek model have been studied to understand various features of different model systems\cite{TonerTu1998,Chateprl2004, Chatepre2008, SudiptaJPCOM, Ihlepre2014}. \\
In these studies, the authors mainly consider a collection of SPPs in a homogeneous system or medium. Recently, there is a growing interest in understanding the effects and advantages of different kinds of inhomogeneity that are omnipresent in nature. Many studies show that inhomogeneity can destroy the LRO present in a clean system 
\cite{Morin2017, Chepizhko2013, Yllanes2017, Quint2015, Sandor2017, Reichhardt2017, Rakesh2018, Toner2018E, Toner2018L}, whereas a few studies discuss special kinds of inhomogeneity that enhance the ordering in the system \cite{RDas2020, SudiptaIS}.
Therefore inhomogeneity can be useful for many practical applications, e.g., crowd control and faster evacuation, etc.\cite{Dorso2011, ZuriguelJSM, Zuriguel2016,Jay,SK, Zuriguel2011,Lin2018,Beat2017,Biplabtopo,SudiptamodelB,Cates2014,Puri2009,Bray1994,bishopprl,MKumar,Zh,Sameer}. \\
In the Vicsek model, each individual interacts through a short-range alignment interaction, and the strength of the interaction is the same for all the particles. But in natural systems, each particle can have a different ability to influence its neighbours based on their individual intelligence or physical strength, etc. However, scientists have not paid much attention to understand the effects of different interaction strengths in a polar flock. 
In a recent study, Bialek {\em et al.} \cite{WilliamPNAS} show that the varying interaction strength of the SPPs results in maximum entropy. Hence, more information transfer among the particles \cite{WilliamPNAS}\cite{sudiptapre}. Surprisingly, in this work, we note that the presence of inhomogeneity in the form of the particles' 
different interaction ability, the system approaches towards a more homogeneous state near to the point of order-disorder transition. More importantly, the flock's response is faster for higher disorder in the interaction strength among the SPPs because each SPP neighbour 
is updated more frequently, which leads to faster information transfer within the flock. 
We also calculate the systems' information entropy \cite{Andrea, Iva, SHANNON,Information} for different disorders and find that the larger the disorder, the more is the systems' information entropy.\\
Further, we have characterised the effect of bond disorder on the nature of disorder-to-order phase transition in the system. The nature of phase transition in an active system has been a matter of great interest in many previous studies \cite{Chatepre2008,SudiptaJPCOM,jpsingh,Mate}. The effect of random impurities shows interesting results in many equilibrium systems \cite{BS,Bs,JM,Janos,Korsh} as well. In the study, M. Durve {\em et al.} \cite{Mihir} have found a first-order phase transition by tuning particles' view angle in the modified Vicsek model.  However, the results observed for phase transitions are very much  model-dependent. 
Our numerical simulation suggests that the nature of the disorder-to-order phase
transition changes from discontinuous to continuous
type by tuning the strength of bond disorder. Also, the system shows the enhanced ordering near the transition point for the larger disorder. \\ 
The rest of the paper is organized as follows. In Sec.\ref{model} ,
we discuss the model and simulation details. In Sec.\ref{results} ,
the results from the numerical simulations are discussed. In Sec.\ref{Discussion} , we conclude the paper with a summary and
discussion on the obtained results.
%in many equilibrium systems  disorder can lead  the transition to discontinuous type \cite{Janos,Korsh}.}} 

\section{Model \label{model}}
{{We consider a collection of $N$ number of polar self-propelled particles (SPPs) moving on a two-dimensional
substrate. SPPs interact through a short-range alignment interaction within a small interaction
radius $R_I$\cite{VicsekT,Chateprl2004,Chatepre2008}. 
Moreover, the strength of interaction of each SPP with its neighbours is {\em different}, unlike the Vicsek model \cite{VicsekT}
of uniform interaction strength. Each SPP is defined by its position 
${\bf r}_i(t)$ with orientation $\theta_{i}(t)$, and it moves
along its direction vector ${\bf n}_{i}(t) = \big {(}\cos(\theta_i(t)), \sin(\theta_i(t)) \big {)}$ with a fixed speed $v_0$. 
The two update equations for the position ${\bf r}_i(t)$ and 
the direction vector ${\bf n}_{i}(t)$
are given by, 
\begin{equation}
	{\bf {r}}_{i} (t + \Delta{t}) = {\bf {r}}_i(t) + {v_{0}}{\bf n}_{i}(t){\Delta{t}}  
\label{eq1}
\end{equation}
\begin{equation}
	{\bf n}_{i}(t+\Delta{t})=\frac {{}\sum _{{j\in R_{I}}}J_{j}{\bf n}_{j}(t)+\eta N_{i}(t){\bf {\bf{{\bf{k}}}}}_{i}(t)}{w_i{(t)}}	
\label{eq2}
\end{equation}
The first equation represents the particle's motion due to its self-propelled nature 
 along the direction vector ${\bf n}_i(t)$ with fixed speed $v_0$. $\Delta t =1.0$ is the unit time step. 
The first term in Eq.(\ref{eq2}) represents the short-range alignment interaction of the $i^{th}$ particle with its neighbours within the interaction radius ($R_{I}$),  
and $J_{j}$ is the interaction strength of the $j^{th}$ neighbour.
The probability distribution of the interaction strength 
$J$, $P(J)$ is obtained from a uniform distribution of
range $[1-\frac{\epsilon}{2}:1+\frac{\epsilon}{2}]$ \cite{MKumar}, 
where $\epsilon$ measures the degree of disorder. 
$\epsilon=0$ corresponds to the uniform
interaction strength $(J_{i}=1$, for all the particles) 
like the Vicsek model \cite{VicsekT}, whereas 
$\epsilon=2$ corresponds to the maximum 
disorder in the system. Furthermore, the second term in Eq.(\ref{eq2}) denotes the vector noise, which measures the particle's error while following its neighbours. 
${\bf k}_i(t)$ is a random unit vector where 
$N_{i}(t)$ denotes the number of neighbours within 
the interaction radius of the $i^{th}$ particle at time $t$.
$\eta$ represents the strength of the noise and can vary from
$0.0$ to $1.0$. $w_{i}(t)$ is the normalization factor, which reduces the right-hand side of the Eq.(\ref{eq2}) to a unit vector. \\
For zero self-propulsion speed, the model reduces to the {\em equilibrium} random bond $XY$-model \cite{Vik, MKumar}.
However, for $\epsilon=0$, the model reduces to the {\em clean} polar flock.
We numerically update the  Eqs.(\ref{eq1}) and (\ref{eq2}) 
for all SPPs sequentially. One simulation step 
is counted after the update of Eqs.(\ref{eq1}) and (\ref{eq2}) once for
all the particles. Periodic boundary condition (PBC) 
is used for the system. All the lengths are measured in terms of interaction radius $R_I=1.0$. The size of the system $L$ is varied from $90$ to $200$. Here system size L means L times the interaction radius. The number density of the system is defined as $\rho _N = \frac{N}{L \times L}$.
We fix the density at $\rho _N = 1.0$ and self-propulsion speed $v_0 = 0.5$. Since for the same density for the clean polar flock, the critical noise is close to $\eta \sim 0.6$. Hence, we limit our study near the critical point, and the noise strength is varied from  $\eta=0.4$ - $0.8$ to study the phase transition, and keeping fixed $\eta=0.62$ to characterise the properties of polar flock near to critical point. 
{{We considered time up to $10^6$ simulation steps and $20$ independent realisations for different values of disorder strength $\epsilon$}}. Although the time is taken to reach the steady state depends on the disorder strength, but for all disorders, steady-state is obtained by $10^4$ simulation steps, and the remaining time is used for time averaging.
\section{Results\label{results}}
\subsection{Disorder-to-order transition\label{Disorder-to-order transition}}
{{First, we study the disorder-to-order transition in the system for different disorder strengths $\epsilon$. Ordering in the system is
characterised by the mean orientation order parameter,
\begin{equation}
	\psi(t)=\frac{1}{N}\vert{\sum^N_{i=1}}n_i(t)\vert	 
\label{eq3}
\end{equation}

\begin{figure}[ht]
\centering
\includegraphics[width=1.0\linewidth]{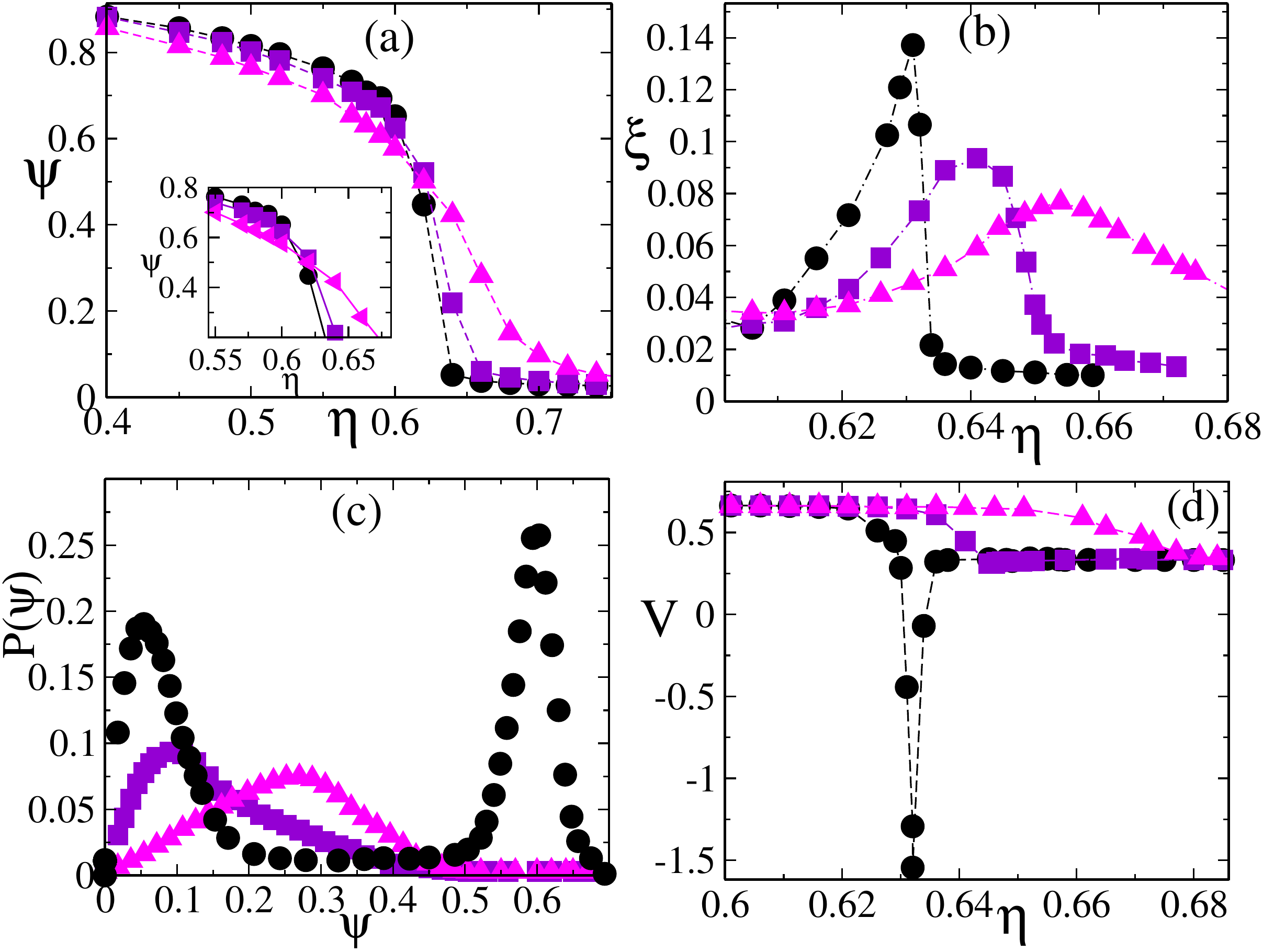}
	\caption{(color online) (a) The plot of the mean orientation order parameter $\psi$  {\em vs.} noise strength $\eta$, {\it inset}: 
	zoomed plot shows enhanced ordering on increasing $\epsilon$. (b) Variation of susceptibility $\xi$ {\em vs.} $\epsilon$. 
	(c) The probability distribution function of order parameter $P(\psi)$ {\em vs.} $\psi$ at the transition point ($\eta_c(\epsilon)$ = 0.625, 0.640, and 0.654 for $\epsilon=0$, 1 and $2$, respectively. (d) Variation of fourth-order Binder cumulant $V$ {\em vs.} $\eta$. 
	Different symbols imply different values of disorder strength $\epsilon = 0$($\circ$), $1.0$($\square$), $2.0$($\triangle$) for the system size L=150 and the density $\rho_N = 1.0$.}
\label{fig:fig1}
\end{figure} 
\begin{figure}[ht]
\centering
\includegraphics[width=1.0\linewidth]{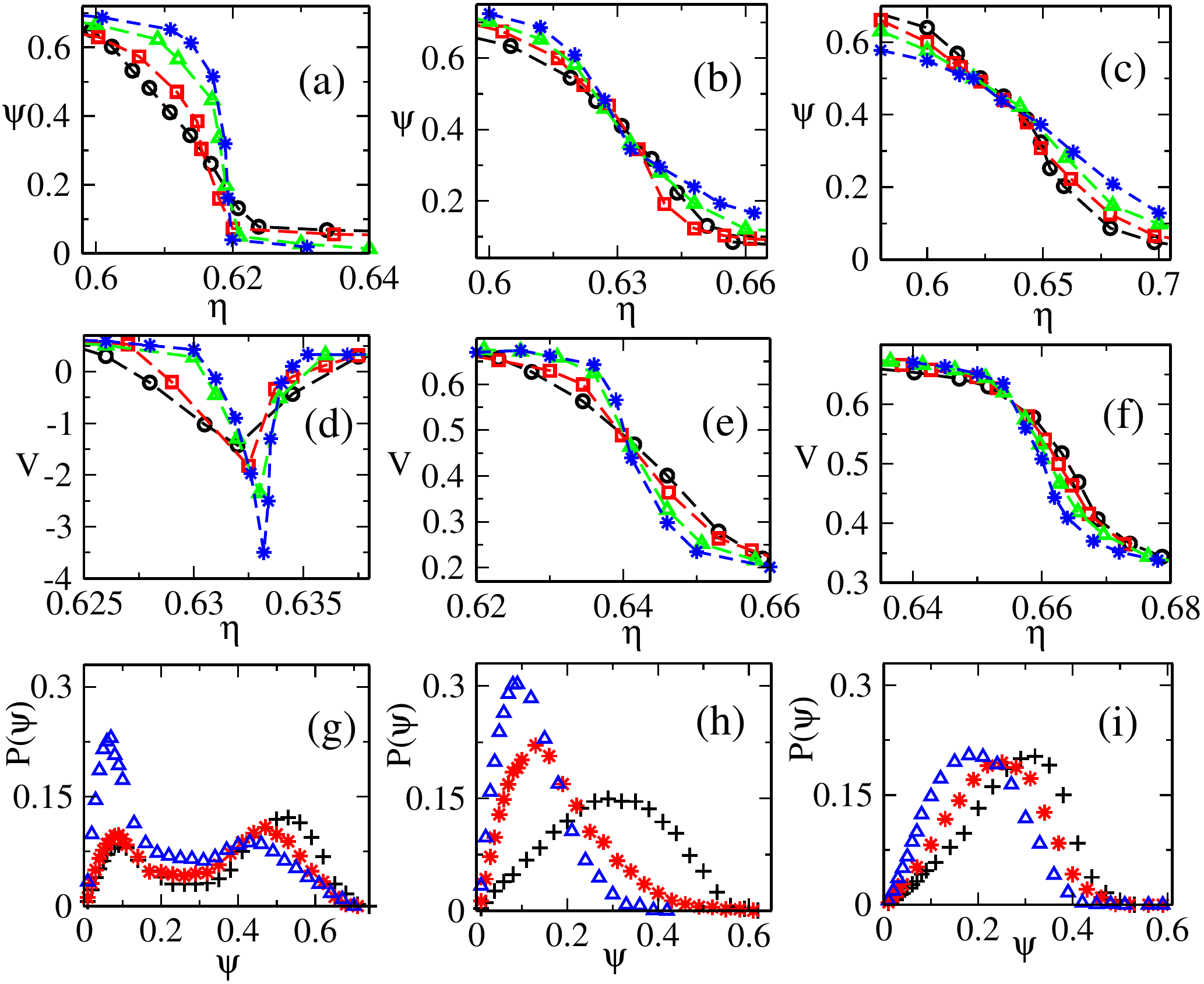}
	\caption{(color online) {{ Plot (a)-(c) show orientation order parameter $\psi$ vs. $\eta$ and (d-f) represent Binder cumulant V {\em vs.} $\eta$ for $\epsilon=0$, 1 and 2. In plots (a)-(f) the black $\circ$, red $\square$, green $\triangle$ and blue $\star$ are for system sizes $L = 90$, $120$, $150$ and $200$, respectively.
(g)-(i) are plots of orientation order parameter distribution $P(\psi)$ keeping system size $L=120$ for three different $\eta$ values. (g) for $\epsilon=0$, $\eta=0.6245,$ 0.6260 and 0.6275, (h) $\epsilon=1$, $\eta=0.6375,$ 0.6390 and 0.6420, and (i) $\epsilon=2$ for $\eta=0.6490,$ 0.6520 and 0.6540, respectively. Symbols with color black `+', red `$\star$' and blue `$\bigtriangleup$' show the increase in $\eta$ values with their respective $\epsilon$ values. In all the cases system density is fixed $\rho_N=1.0$}}.}
\label{fig:fig2}
\end{figure} 
In the ordered state, i.e., when most particles are moving in the same direction, then $\psi$ will be closer to 1.0
and of the order of $\frac{1}{\sqrt{N}}$ for a random disordered state. In Fig.\ref{fig:fig1}(a), we have shown the variation of $\psi(t)$  with the
noise strength $\eta$ for three different $\epsilon$ $=(0,$ 1 and 2). For $\epsilon=0$,  the variation of
$\psi$ shows a sharp change from $\psi  \sim 1.0$ to $\sim 0.0$. This kind of change is a common feature of first-order phase  
transition \cite{Chatepre2008,SudiptaJPCOM,Ihlepre2014,Morin2017,Chepizhko2013}. 
Whereas for $\epsilon=2$, $\psi$ varies continuously, and the transition has a signature of second-order phase transition. The variation of $\psi$ for $\epsilon=1$ shows the intermediate behaviour. The plot of order parameter fluctuation or the susceptibility 
$\xi=\sqrt{<\psi^2>-<\psi>^2}$ is shown in Fig.\ref{fig:fig1}(b) where $<...>$ denotes the average over
steady-state time. The critical noise $\eta_c$ is obtained from the maximum of $\xi$. The $\eta_c(\epsilon)$ shifts towards the right on increasing $\epsilon=0,$ 1 and 2, respectively.
To understand further the nature of the phase transition, 
we plot order parameter probability distribution function (PDF)  $P(\psi)$ vs. $\psi$ in Fig.\ref{fig:fig1}(c) at the critical noise $\eta_c(\epsilon) = 0.625,$ 0.640 and 0.654 
for three $\epsilon=$  $0,$ 1 and $2$, respectively. 
For $\epsilon=0$, there is a clear bimodal nature of $P(\psi)$ which gradually changes to unimodal on increasing $\epsilon$.  
To characterise further, the nature of the transition  for $\epsilon$ $=(0,$ 1 and 2) in Fig.\ref{fig:fig1}(d), we calculate the fourth-order cumulant or the Binder 
cumulant $V=1-\frac{<\psi^4>}{3<\psi^2>^2}$ vs. $\eta$.
\begin{figure}[ht]
\centering
\includegraphics[width=1.0\linewidth]{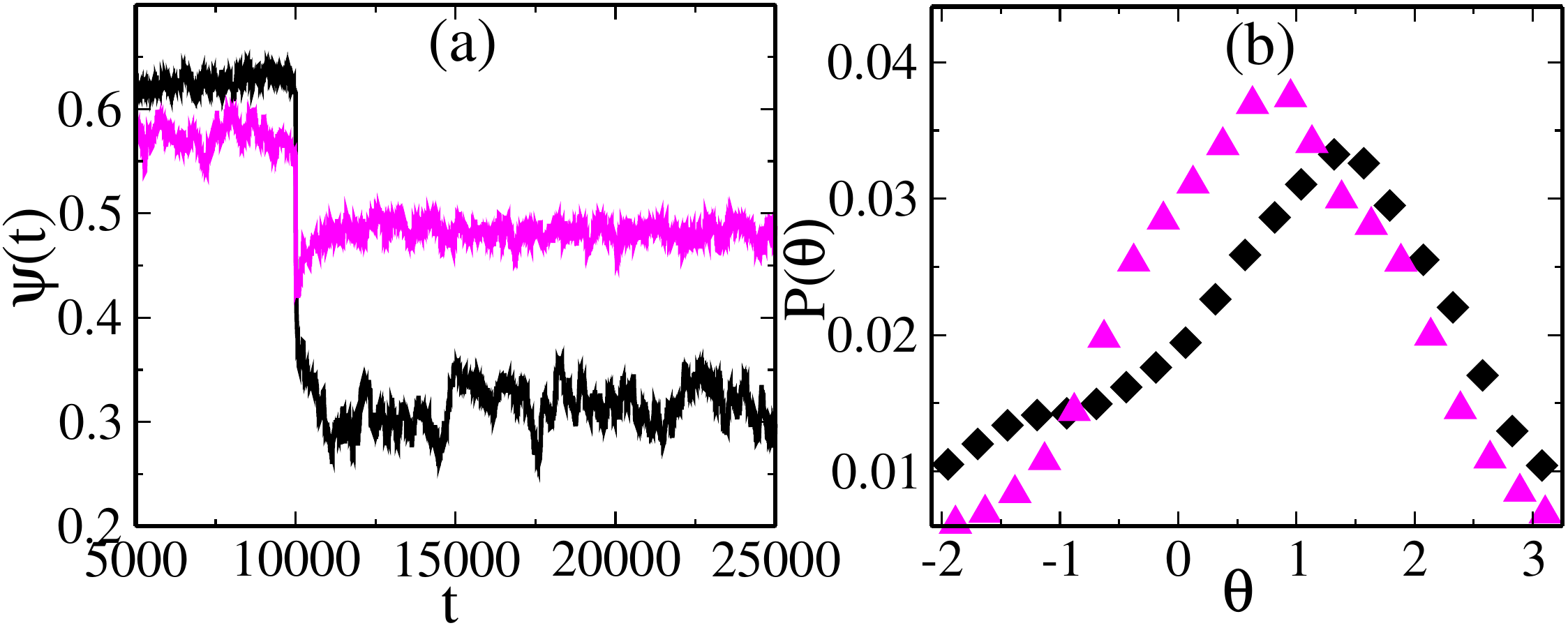}
\caption{(color online) (a) Time series of $\psi(t)$ for disorder $\epsilon =1.0$ (black line) and $\epsilon=2.0$ (magenta line). The two types of quenched impurities 
	with orientations $\pm \frac{\pi}{2}$ are introduced at time $t=10,000$. (b) PDF for the orieantation distribution $P(\theta)$ {\em vs.} mean orieantation $\theta$, 
	for $\epsilon = 1.0$ ($\square$) and $\epsilon = 2.0$ ($\triangle$). All the plots are for system size $L=100$, noise strength $\eta = 0.62$ and the density $\rho_N = 1.0$.}
\label{fig:fig3}
\end{figure}  
We plot $V(\eta)$ vs. $\eta$ in Fig.\ref{fig:fig1}(d). It shows strong discontinuity between $V=1/3$ (for disordered state) to
$V=2/3$ (for ordered state) as we approach critical $\eta_c$ for $\epsilon=0$. However, it goes smoothly between a disordered state $(V=1/3)$ to an ordered state $(V=2/3)$ for $\epsilon=2$.
\begin{figure}[ht]
\centering
\includegraphics[width=1.0\linewidth]{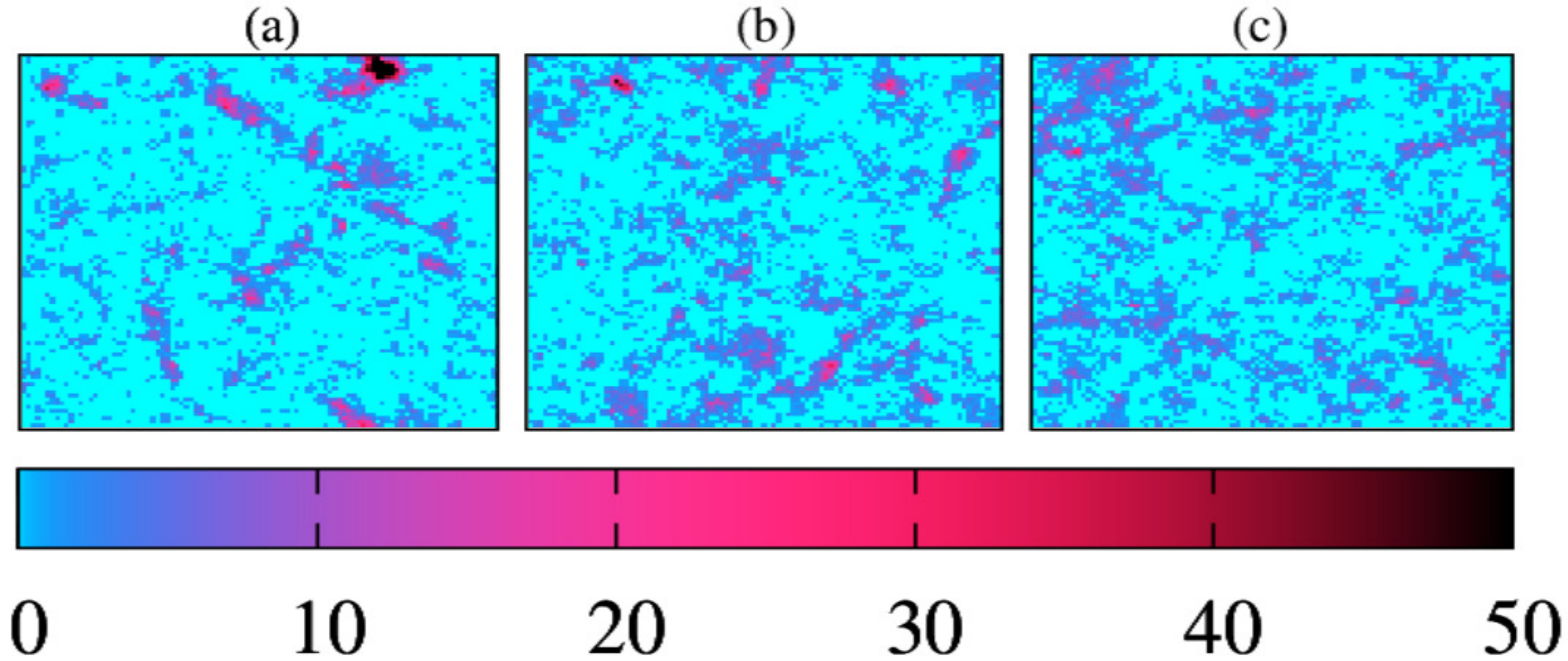}
\caption{(color online) Plot (a), (b) and (c) are the real space snapshots  for three $\epsilon=0$, 1 and 2, respectively. {{The color bar shows the number of particles in the subshells. All the parameters are the same as in Fig.\ref{fig:fig3}.}} }
\label{fig:fig4}
\end{figure}\\
{\bf{\em Finite-size analysis}:}- {{To characterise further, the nature of the phase transition more  precisely, we perform the finite-size analysis of the system for different strengths of disorder $\epsilon$. First, we plot the global orientation order parameter $\psi$ vs. noise strength $\eta$ for three disorder strengths $\epsilon=0,$ 1 and 2 in Fig.\ref{fig:fig2} (a)-(c), for four system sizes $L=90$, $120$, $150$ and $200$, respectively. In Fig.\ref{fig:fig2}(a) for $\epsilon=0$, with the increase in the system size, clearly, the value of order parameter $\psi$, becomes more discontinuous on increasing system size. In Fig.\ref{fig:fig2}(b) for $\epsilon=1$, $\psi$ becomes more continuous on increasing system size. 
For $\epsilon=2$ as shown in Fig.\ref{fig:fig2}(c), the $\psi$ curves become 
more continuous on increasing system size compared to former $\epsilon$ values. 
Moreover, to further understand the effect of finite size on the phase transition, we calculate fourth-order Binder cumulant $V$ for $\epsilon=0,$ 1 and 2 in Fig.\ref{fig:fig2}(d)-(f) for four system sizes $L=90, 120$, 150 and 200. For $\epsilon=0$ in Fig.\ref{fig:fig2}(d), with the increase in the system size, $V$ changes sharply, showing discontinuous phase transition. While for $\epsilon=1$ in Fig.\ref{fig:fig2}(e), V shows a smooth crossover. Further, in Fig.\ref{fig:fig2}(f) for $\epsilon=2$, we get a clear crossing of $V$ at a single point with respect to different system sizes. This is one of the clear signatures of continuous transition{{\cite{finite}}}.
Finally, we calculate the orientation order parameter probability distribution function (PDF) $P(\psi)$ for $\epsilon=0,$ 1 and 2 in Fig. \ref{fig:fig2}(g)-(i), keeping system size $120$ for three different $\eta$'s. In Fig.\ref{fig:fig2}(g), for $\epsilon=0$, there is a clear bimodal signature for three $\eta=0.6245$, 0.6260 and 0.6275. In Fig.\ref{fig:fig2}(h) for $\epsilon=1$ for three $\eta=0.6375$, 0.6390 and 0.6420. Here we again find unimodal nature, but with a large tail. Further, in Fig.\ref{fig:fig2}(i) for $\epsilon=2$, $P(\psi)$ for three $\eta=0.6490,$ 0.6520 and 0.6540. Here it is very clear that $P(\psi)$ shifts towards the left slowly (continuous manner) with increased $\eta$, which confirms there is a clear continuous nature of the phase transition.}}
Hence the above finite-size analysis and behaviour of PDF near critical point suggest the change in the nature of the phase transition from discontinuous type to the continuous type on the increasing strength of disorder $\epsilon$.
{{Further, we characterise the enhanced ordering near to the critical values of $\eta$ and shift of $\eta_c(\epsilon)$ towards higher values for $\epsilon=0$, 1 and 2 as shown in the inset of Fig.\ref{fig:fig1} (a)}}.\\
{\bf{\em Enhanced ordering}} :-
To understand the enhanced ordering mechanism, we perform a small perturbative study on the system. 
Since we find enhanced ordering  near  $\eta \sim 0.6$, the perturbation is imposed at $\eta=0.62$ for 
finite disorder  $\epsilon=1$ and $2$.\\
In the perturbative study, the system is awaited to reach the
steady-state ($t = 10^4$) and once the steady-state is
reached; 
we randomly choose $5\%$ of the particles  and out of which the direction of $2.5\%$ particles with  $J>1$ quench to 
the direction $\frac{\pi}{2}$ and remaining $2.5\%$ with $J<1$ are quenched to the direction $\frac{-\pi}{2}$.
Once this perturbation is applied, the system will respond to it and mean order 
parameter $\psi(t)$ shows a dip and then relaxes to a new steady-state with a relatively lower value of $\psi(t)$ as shown in Fig.\ref{fig:fig3}(a). Very clearly, before perturbation, $\psi$ is lower for $\epsilon =2$, hence a more ordered state for the lower disorder. But after perturbation, which is selectively
for particles with higher and lower $J$ values, the response is different for $\epsilon=1$ and $2$. For $\epsilon=2$, after perturbation $\psi$ is larger 
compared to $\epsilon=1$. Hence more ordered state for the larger disorder. In the plot of Fig.\ref{fig:fig3}(b), we plot the orientation probability distribution  function (PDF)   $P(\theta)$ of the orientation of the particles $\theta$. 
For $\epsilon=1$, the $P(\theta)$ shows two distinct peaks for $\theta=\pm \pi/2$, but the peak for $\pi/2$ or response to higher $J$ is more. For $\epsilon=2$, the mean of the  
$P(\theta)$ shifts towards the non-zero $\theta$, hence the system's response happens globally, and the whole system is polarised in the direction of quenched particles with larger $J$ values. \\
Now, we further study the consequence of such enhanced ordering for larger disorder on the polar flock.
Moreover, this dominated alignment is responsible for the shifting of transition point $\eta_c$ towards higher values.\\
\subsection{Properties of the polar flock\label{Properties of polar flock}}
\begin{figure}[ht]
\centering
\includegraphics[width=1.0\linewidth]{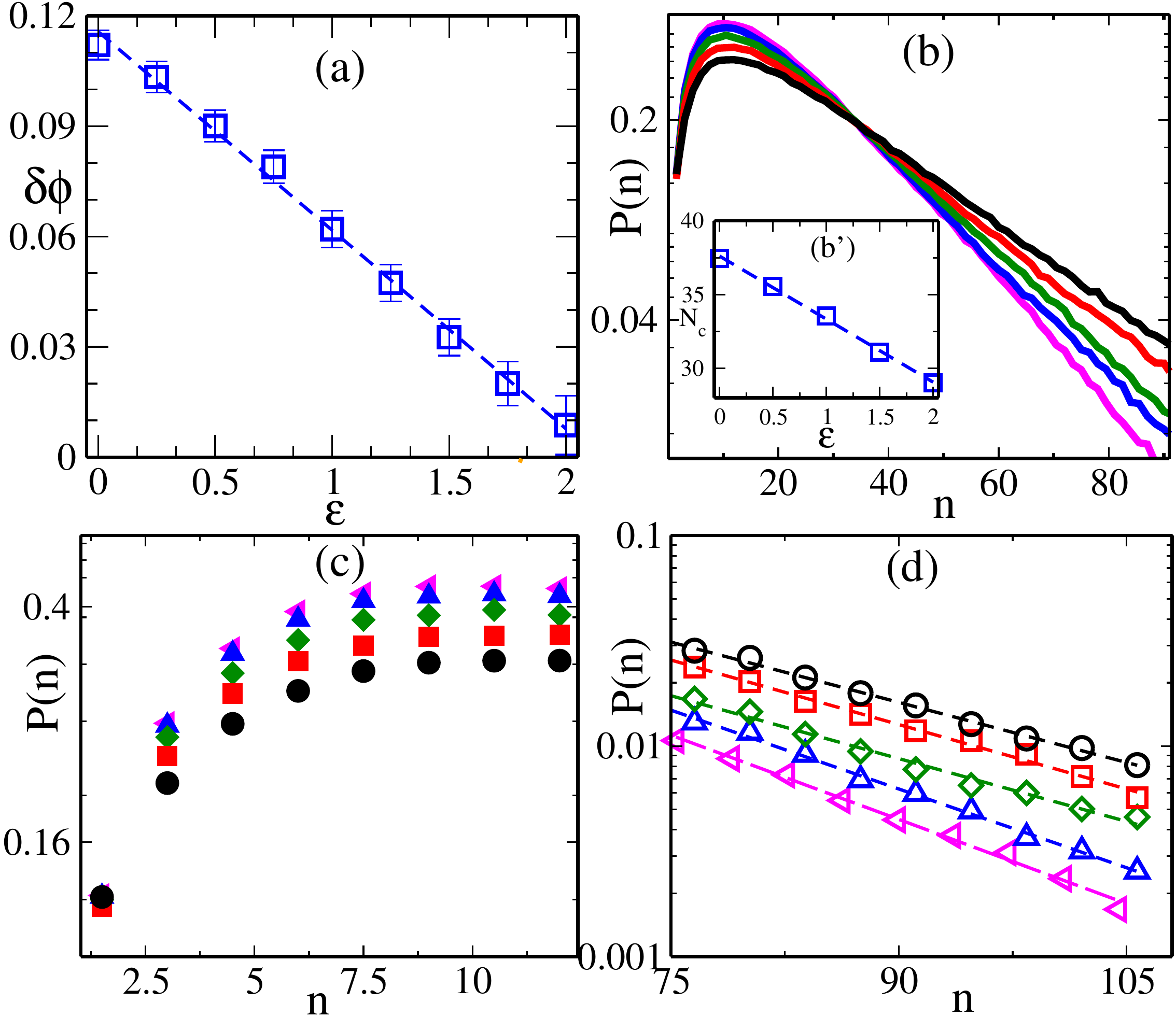}
\caption{(color online) (a) Plot of density phase separation order parameter $\delta{\phi}$ vs. $\epsilon$  with blue squares. The blue dotted line shows the linear decay of $\delta{\phi}$. (b) P(n) vs. n for $\epsilon=0,$ 0.5, 1, 1.5 and 2. Different colors with lines black, red, green, blue and magenta are for $\epsilon=0,$ 0.5, 1.0, 1.5 and 2.0. The inset of Fig.(b) shows the mean number of particles $N_c$ vs. $\epsilon$ where the blue dotted line shows linear decay of $N_c$. (c) and (d) show the zoom plot of (b) near to the head and tail where the tail parts are fitted with exponential function with dotted lines. Symbols with black (circles), red (squares), green (diamonds), blue(triangles up) and magenta (triangles left) colors are for $\epsilon=0,$ 0.5, 1.0, 1.5 and 2.0 in (c) and (d). Also, (b), (c) and (d) are in semi-log y-axis. All the parameters are the same as in Fig.\ref{fig:fig3}.}
\label{fig:fig5}
\end{figure}
How does disorder affect the density fluctuations in the system?
We plot real space snapshots of the local density (calculated in a small region of unit size square sub-shell) in Fig.\ref{fig:fig4}(a)-(c) for three values of $\epsilon=(0,$ 1 and $2)$ at time $t=10^6$. 
For clean polar flock ($\epsilon=0$), particles form isolated clusters. Whereas with a non-zero $\epsilon$, these isolated clusters break, and the system gets into a more homogeneous state.
  To further confirm this, we calculate the density phase separation order parameter,
  $\delta\phi$ vs. $\epsilon$ (where $\delta \phi(\epsilon)$ is the deviation of the number of particles among the sub-cells), as shown in Fig.\ref{fig:fig5}(a). 
  We calculate $\delta\phi$ by dividing the whole $L\times{L}$ system into unit sized sub-cells, 
  $\delta\phi(\epsilon)=\sqrt{\frac{1}{L^2}{\sum^{L^2}_{j=1}}{(\phi_j(\epsilon))^2}-(\frac{1}{L^2}{\sum^{L^2}_{j=1}}{\phi_j(\epsilon)})^2}$ 
  where $\phi_j$ is the number of particles in the $j^{th}$ sub-cell and $\langle....\rangle$ represents averaging over $20$ realisations. 
  We note that $\delta\phi$  decreases in a linear fashion with increasing $\epsilon$ as shown in Fig. \ref{fig:fig5}(a). 
  Hence system becomes more homogeneous with increasing the random bond disorder $\epsilon$ in the system. 
  Furthermore,  in Fig.\ref{fig:fig5}(b), we plot the probability  distribution function (PDF) of a number of neighbours $P(n)$ for different 
 values of $\epsilon=0$, 0.5, 1, 1.5 and $2$, respectively. Sharper tail for large $n$ for the higher value of $\epsilon=2$ shows that the system is approaching 
 towards a more homogeneous
 state or clusters of smaller size while the longer tail, for lower values of $\epsilon=0$, hence bigger clusters. In the inset of Fig.\ref{fig:fig5}(b), 
 we plot the mean number of particles $N_c$ with $\epsilon$ where $N_c$ is obtained by fitting the tail of the main plot by the exponential function $\exp({\frac{-n}{N_c}})$. 
 This shows that $N_c$ decreases linearly with an increase in the value of $\epsilon$. 
 Similarly, when zoomed for smaller $n$ as shown in Fig.\ref{fig:fig5}(c), $P(n)$ for larger $\epsilon$ is higher as compared to smaller $\epsilon$. Hence formation of small clusters have more probability for the larger disorder. Fig.\ref{fig:fig5}(d) shows the zoomed tail of the $P(n)$. 
 \begin{figure}[ht]
\centering
\includegraphics[width=1.0\linewidth]{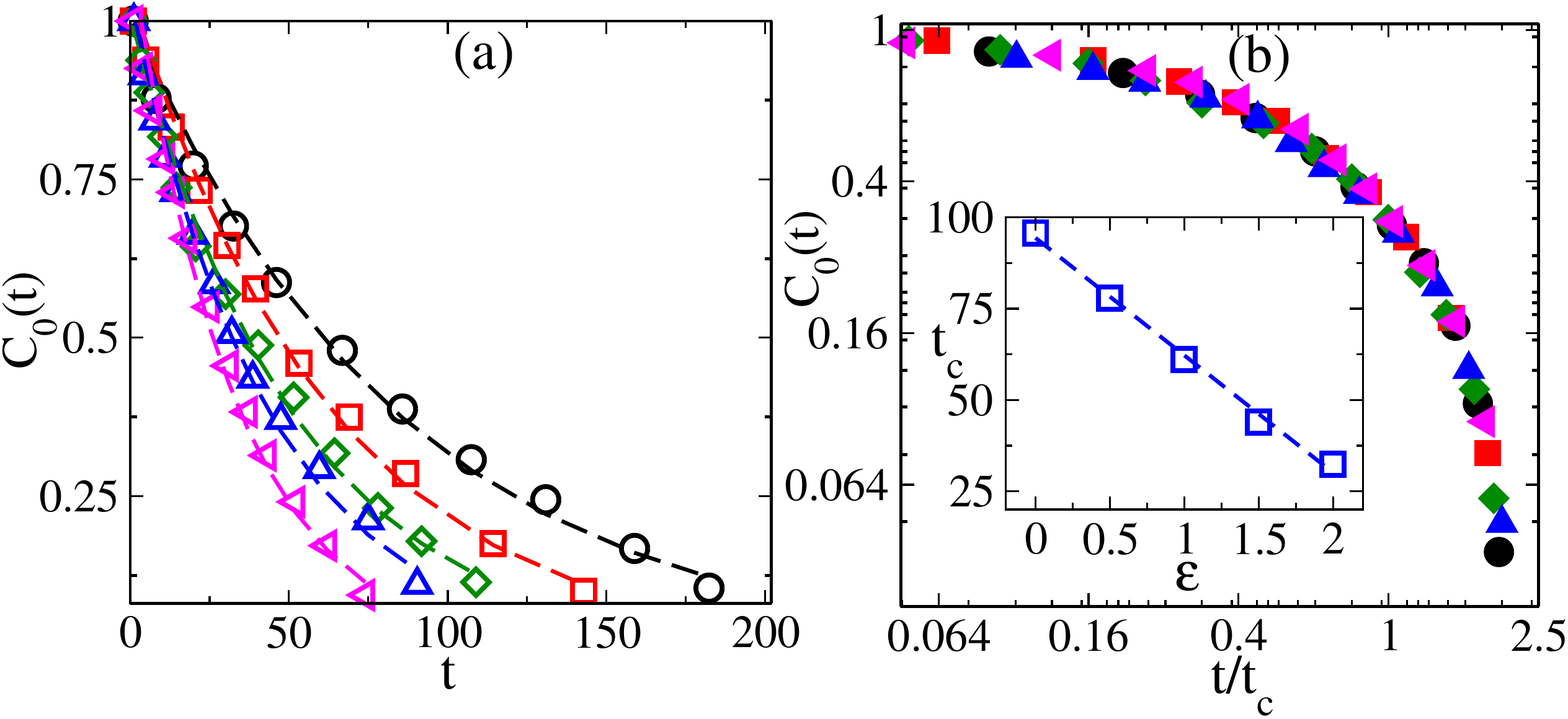}
\caption{(color online) (a) Plot of OACF $C_0(r,t)$ {\em vs.} $t$. for $\epsilon$ = $0.0$ (circles), $0.5$ (squares), $1.0$ (diamonds), $1.5$ (triangles up) and $2.0$ (triangles left). Dashed lines are fit to exponential to the data (symbols). (b) Plot for $C_0(r,t)$ {\em vs.} scaled time  $t/t_c$; and $t_c$ {\em vs.} $\epsilon$ (inset) where the dashed lines are a linear fit to the data. All other parameters are the same as in Fig. \ref{fig:fig3}}
\label{fig:fig6}
\end{figure}
\subsection{Accelerated response to external perturbation \label{Random bond disorder promotes faster information transfer}}
We claim that enhanced ordering near-critical regions, and homogeneous density clusters promote faster response among the flock. 
To confirm the same, we perform another perturbation to the well-ordered flock in the steady-state and calculate its response. 
We randomly select a fraction of particles $1\%$ and quench their direction to a randomly selected fixed orientation. With time all other particles will rotate in that direction. 
Their response to the direction of quench is measured by calculating the orientation  auto-correlation function (OACF) $C_0(t)=\langle{\cos{\theta_i(t)}-\theta_i(0)}\rangle - \langle{\cos{\theta_i(T)}-\theta_i(0)}\rangle$.
Where $\theta_i(t)$ and $\theta_i(0)$ are the orientation of the $i^{th}$ particle at time $t$ and $0$ from the time of quench, and $T$ is the late time when approximately all the particles are oriented in the direction of the quench. $\langle {....} \rangle$ denotes averaging over all the SPPs over $30$ independent realizations. 
In Fig.\ref{fig:fig6}(a), OACF $C_0(t)$ decays exponentially and shows the sharper decay with the increase in the strength of disorder $\epsilon$. Therefore, the response of the flock to external perturbation becomes faster with the increase in $\epsilon$. 
In Fig.\ref{fig:fig6}(b), we plot the $C_0(t)$ {\em vs.} scaled time $t/t_c$, where $t_c$ is obtained from the fitting of $C_0(t) $ to $\exp(-t/t_c)$. The inset of Fig.\ref{fig:fig6}(b) shows the variation of  $t_c$ vs. $\epsilon$. $t_c$ shows linear decay with $\epsilon$, which confirms the faster response of the flock towards external perturbation with an increase in the value of $\epsilon$.

\subsection{Disorder increases the systems' information entropy\label{Disorder increases system's information entropy}}
 \begin{figure}[ht]
\centering
\includegraphics[width=1.420\linewidth]{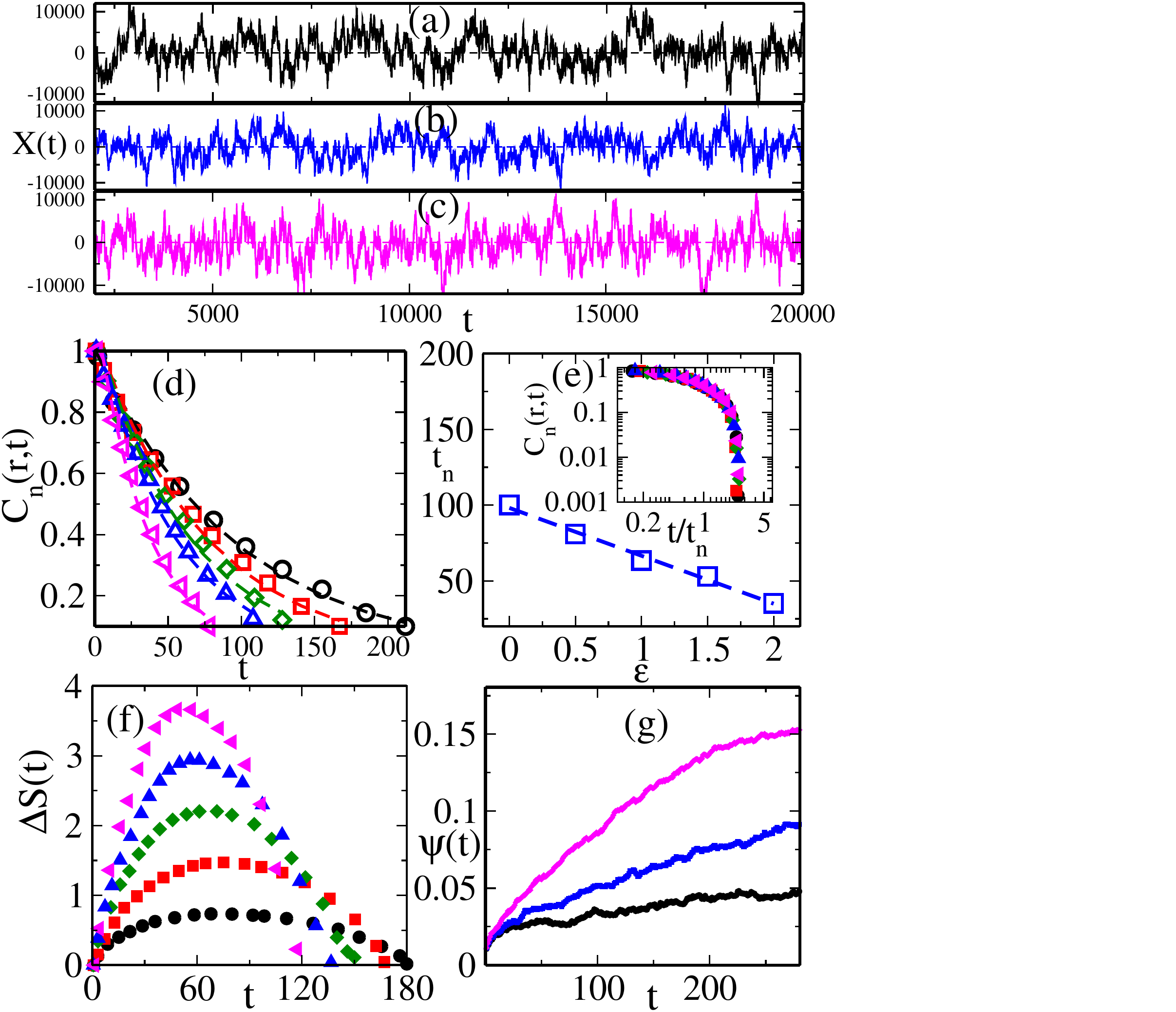}
	 \caption{(color online) (a), (b) and (c) show the variation of $X(t)$ {\em vs.} $t$. Black, blue and magenta colors are for $\epsilon=0,$ 1 and $2$, respectively. (d) 
	 Variation of neighbour fluctuation autocorrelation $C_n(r,t)$ {\em vs.} $t.$ Symbols with black (circles), red (squares), green (diamonds), blue (triangles up) and magenta (triangles left) colors are for $\epsilon=0,$ 0.5, 1.0, 1.5 and $2.0$, respectively. Dashed lines are fit to exponential. (e) The plot of $t_n$ vs. $\epsilon$  shows linear decay with $\epsilon$; inset; the plot of correlation $C_n(r,t)$ {\em vs.} scaled time $t/t_n$. (f) The plot of the systems' information entropy $\Delta{S}$(t) {\em vs.} $t$. (g) Time evolution of  $\psi(t)$ with time $t$ where colors black, blue and magenta are for $\epsilon=0,$ 1 and $2$, respectively. All other parameters are the same as in Fig. \ref{fig:fig3}}
\label{fig:fig7}
\end{figure}
Further, we claim that the accelerated response to external perturbation is due to neighbours' frequent updates for high disorder strength. \\
We define the update in the
 neighbour list of the SPPs as $X(t)=\frac{1}{N}\sum^{N}_{i=1}((<N_R^i(t)\times N/2>)-\sum_{j\in R}j)$. Where $N_R^i$ is the number of SPPs inside the interaction radius of the $i^{th}$ particle, $N$ is the total number of particles in the system and the second term on the right-hand side is the sum over all the particle indices $j$ inside the interaction radius of the $i^{th}$ particle.  
The time  series  of $X(t)$  oscillates  around  $0$  for  different  values of $\epsilon$,  as  shown   in  Fig.\ref{fig:fig7}(a), (b) and (c).  The frequency of oscillation of $X(t)$  increases with increasing $\epsilon$.  The increase in the oscillation frequency of $X(t)$  suggests more frequent updates of the neighbour list and the decrease in the magnitude of $X(t)$ implies a lesser number of neighbours inside the interaction radius of an SPP. Furthermore, we calculate the neighbour autocorrelation function, 
\begin{equation}
 C_n(t)=\big\langle\frac{\sum^{T-t}_{t^{'}=1}(X(t^{'})-\overline{X})(X(t^{'}+t)-\overline{X})}{\sum^{T}_{t^{'}=1}(X(t^{'})-\overline{X})^2}\rangle
 \label{eq4} 
 \end{equation}
where $\overline{X}$ is the mean value of $X(t)$ over the total time $T$ and $t < T$. $\langle ....\rangle$ represents averaging over $20$ independent realisations. In Fig.\ref{fig:fig7}(d), faster decay of $C_n(t)$ with increase in the disorder strength $\epsilon$, suggests more frequent update of neighour list. Also, in the inset of Fig.\ref{fig:fig7}(e), we plot the scaled correlation $C_n(t)$ vs. $t/t_n$ where $t_n$ is obtained by fitting the exponential function to $\exp(\frac{-t}{t_n})$. 
In Fig.\ref{fig:fig7}(e), we have shown the variation of  $t_n$ with $\epsilon$ which decays linearly. 
Now we use the systems' information entropy \cite{Information, Andrea} approach to show that the larger the disorder, the larger is the systems' information entropy and hence the more information transfer among the SPPs. The faster information transfer in more disorder system is due to the possibility of more number of accessible states for the particles. Each state can be defined as the new neighbour in the chosen particle's contact list, if we denote $P_s$, as the probability of being in the $i^{th}$ states from the set of all possible accessible states. 
If a particle changes its neighbours frequently, it is exploring more number of neighbouring particles and hence more number of states. Hence the neighbour autocorrelation, $C_{n,s}(t)$, 
( the quantity inside the $< ...>$ of Eq.(\ref{eq4}) is the neighbour autocorrelation for one state. The subscript $s$ denotes the different independent configurations and hence
different sets can be generalised as different independent configurations. 
And $C_{n,s}(t)$ and $P_s(t)$ are equivalent. One is the measure of the probability of being in a given neighbour list, and is the same as $P_s(t)$.
As time progresses $C_{n,s}(t)$ decreases, and hence more and more states are accessed. Hence we  define the systems' information entropy of the system as  $\Delta S(t) = -\sum_{s} C_{n,s}(t) \ln_2 C_{n,s}(t)$, where summation $s$ is over all possible realisations. Larger the systems' information entropy, larger the available microstate for the particles, and hence the more information transfer among the flocks.

\begin{figure}[ht]
\centering
\includegraphics[width=0.90\linewidth]{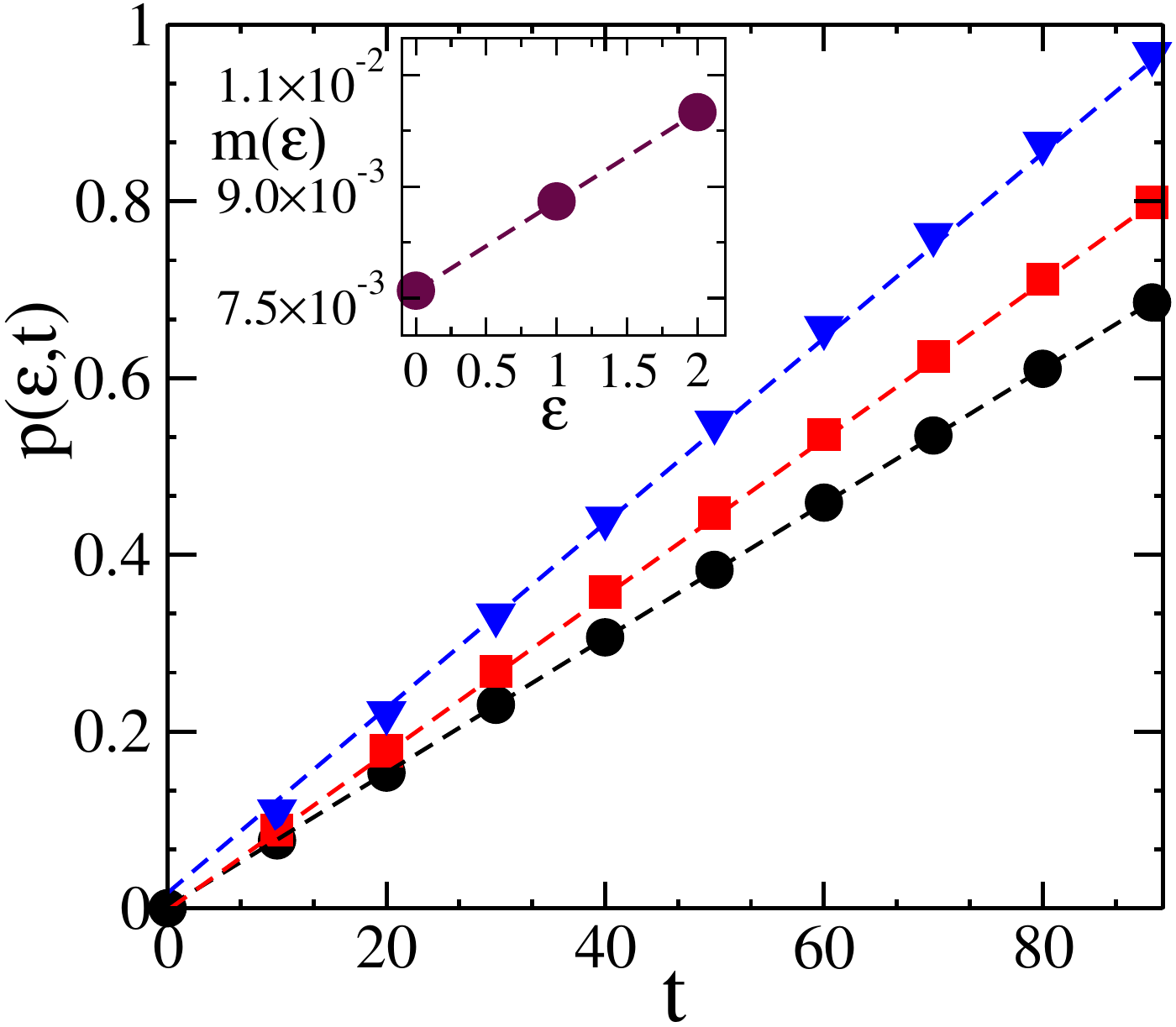}
\caption{(color online) {{The plot  shows the probability of newly visited particles $p(\epsilon, t)$ with respect to time t for three values of $\epsilon$. Black $\circ$, red $\square$, and blue $\triangle$ are for $\epsilon=0,$ 1 and 2, respectively. Color dotted lines show the data fit with fitting function $p(\epsilon,t)=m(\epsilon)\cdot t$ where $m(\epsilon)$ is the slope. In the inset, we show the $m$ vs. $\epsilon$ which increases linearly. 
Other parameters are the same as in Fig.\ref{fig:fig3}}}}
\label{fig:fig8}
\end{figure}
We plot the variation of the systems' information entropy $\Delta{S}(t)$  for different disorders in Fig.\ref{fig:fig7}}(f). 
We note that $\Delta{S}(t)$ increases with $\epsilon$ which further confirms that particles are exploring more states for higher disorder strength hence more information transfer.   
Also, in Fig.\ref{fig:fig7}(g), we have plotted the time evolution of order parameter in the early time, which shows that for larger $\epsilon$, the system reaches the ordered state quicker in comparison to the lower $\epsilon$. 
Further, we also show the quicker update of the neighbour list for a  single particle. In Fig.\ref{fig:fig8} , we plot the fraction of new particles $p(\epsilon, t)$ 
in the neighbour list of a given particle from some reference time {{$t_0=0$}}. At time $t_0$, all the particles are labeled as old, hence $p=0$. 
Time in Fig.\ref{fig:fig8} , is measured from the reference time $t_0$, hence reference time is time zero on the $x-$axis. As time progresses, new particles come in the contact list of the given particle, and $p(\epsilon ,t)$ starts to increase. At a very late time, all the old particles are gone out of the neighbour list, and hence $p=1$. As shown in the figure, for the larger disorders, new neighbours are updated faster than for the small disorder. In the inset of Fig.\ref{fig:fig8} , {{we show the linear increase of slope $m(\epsilon)$ with respect to $\epsilon=0$, 1 and 2 where $m(\epsilon)$ is obtained from fitting the main plot with the linear fitting function}}.

\section{Discussion \label{Discussion}}
We introduce a minimal model for a collection of self-propelled particles with bond disorder. 
Each particle has a different ability (interaction strength) to influence its neighbours.
The varying interaction strength is obtained
from a uniform distribution, and it can be varied from $[1-\epsilon/2: 1+\epsilon/2]$, where $\epsilon$ is
the disorder strength. For $\epsilon=0$, the model reduces to the
constant interaction strength model or the Vicsek-like model \cite{VicsekT}. We have studied the steady-state characteristics for different strengths of the disorder near to order-disorder transition. To our surprise, bond disorder leads to faster information transfer within the flock viz; the systems' information entropy gets increased.\\
 Our numerical study also shows that the disorder-to-order transition is discontinuous in the disorder-free system and changes to continuous type with an increase in disorder. Furthermore, the transition point shifts towards the higher $\eta$ for the large disorder.\\
Our study provides a new direction to understand the effect of intrinsic inhomogeneity in many natural active systems. It shows how the bond-disorder 
in the system can enhance ordering, and faster information transfer among the particles. Such properties can be useful for many applications: like
the faster evacuation of active particles, and also for crowd control in many
social gatherings \cite{Jheb, Mspe}

\section{Acknowledgement \label{Acknowledgement}}
SM and SK thank DST-SERB India, ECR/2017/000659 for the financial support.
The support and the resources provided by PARAM Shivay Facility under the National Supercomputing Mission, Government of India at the Indian Institute of Technology, Varanasi are gratefully acknowledged. Computing facility at Indian Institute of Technology(BHU), Varanasi is gratefully acknowledged.
%%%%%%%%%%%%%%%%%%%%%%%%%%%%%%%%%%%%%%%%%%%%%%%%%%%%%%

\end{document}